\documentclass[floatfix,aps,twocolumn,prl,superscriptaddress, amsfonts]{revtex4}

\usepackage{graphicx}
\usepackage{dcolumn}
\usepackage{natbib}
\usepackage{bm}

\begin{document}
\title{Impurity Band Conduction in a High Temperature Ferromagnetic Semiconductor}
\author{K.S. Burch}
\altaffiliation[Permanent address: ]
{Los Alamos National Laboratory, MS G756, MST-CINT, Los Alamos, NM 87545}
\email{kburch@lanl.gov}
\affiliation{Department of Physics, University of California, San Diego, CA 92093-0319}
\author{D.B. Shrekenhamer}
\affiliation{Department of Physics, University of California, San Diego, CA 92093-0319}
\author{E.J. Singley}
\altaffiliation[Permanent address: ]
{Department of Physics, California State University, East Bay, CA 94542}
\affiliation{Department of Physics, University of California, San Diego, CA 92093-0319}
\author{J. Stephens}
\affiliation{Center for Spintronics and Quantum Computation, University of California, Santa Barbara, CA 93106}
\author{B.L. Sheu}
\affiliation{Department of Physics and Materials Research Institute, The Pennsylvania State University, University Park, Pennsylvania 16802}
\author{R.K. Kawakami}
\altaffiliation[Permanent address: ]
{Department of Physics, University of California, Riverside, CA 92521}
\affiliation{Center for Spintronics and Quantum Computation, University of California, Santa Barbara, CA 93106}
\author{P. Schiffer}
\affiliation{Department of Physics and Materials Research Institute, The Pennsylvania State University, University Park, Pennsylvania 16802}
\author{N. Samarth}
\affiliation{Department of Physics and Materials Research Institute, The Pennsylvania State University, University Park, Pennsylvania 16802}
\author{D.D. Awschalom}
\affiliation{Center for Spintronics and Quantum Computation, University of California, Santa Barbara, CA 93106}
\author{D.N. Basov}
\affiliation{Department of Physics, University of California, San Diego, CA 92093-0319}

\begin{abstract}
    The band structure of a prototypical dilute magnetic semiconductor (DMS), Ga$_{1-x}$Mn$_{x}$As, is studied across the phase diagram via optical spectroscopy. We prove that the Fermi energy ($E_{F}$) resides in a Mn induced impurity band (IB). Specifically the changes in the frequency dependent optical conductivity ($\sigma_{1}(\omega)$) with carrier density are only consistent with $E_{F}$ lying in an IB. Furthermore the large effective mass ($m^*$) of the carriers inferred from our analysis of $\sigma_{1}(\omega)$ supports this conclusion. Our findings demonstrate that the metal to insulator transition in this DMS is qualitatively different from other III-V semiconductors doped with non-magnetic impurities. We also provide insights into Ga$_{1-x}$Mn$_{x}$As' anomalous transport properties.\end{abstract}

\maketitle

The dilute magnetic semiconductor Ga$_{1-x}$Mn$_{x}$As presents a unique opportunity to study carrier mediated magnetism in a well controlled environment. It is generally accepted that the ferromagnetic interaction between the local moments provided by the substitional Mn ($Mn_{Ga}$) is mediated by the holes also donated by the $Mn_{Ga}$, and that for $x<0.04$ these carriers reside in a Mn induced impurity band (IB).
\cite{SamarthReview} However the nature of the states at higher carrier densities (\textit{p}), relevant for high ferromagnetic transition temperatures ($T_{C}$), remains controversial. It is often assumed that Ga$_{1-x}$Mn$_{x}$As falls within the Mott picture of the metal to insulator transition (MIT), wherein the IB eventually dissolves into the GaAs valence band (VB). Theoretical studies based on this approach have successfully described some of the properties of Ga$_{1-x}$Mn$_{x}$As.\cite{vbthry,sinova} Others have suggested that the persistence of the Mn induced impurity band (IB) at all carrier densities is critical to describing the physics of Ga$_{1-x}$Mn$_{x}$As.\cite{MillisDasSarma,imthry, Flatte, moreno} Previous studies have supported the notion that the IB exists in the metallic state at low \textit{p}.\cite{Fujimori,kbelip,jasonpapers,Ploog, DFH} Nonetheless this letter is the first to conclusively demonstrate the existence of the IB in samples with reduced compensation and defect concentrations revealing the highest reported values of T$_{C}$ for the concentrations of Mn studied. We also uncover the origin of the small mobility in Ga$_{1-x}$Mn$_{x}$As.

Our determination that $E_{F}$ lies in an IB is enabled by the distinct free-carrier absorption we observe in highly conductive films, as well as careful analysis of the electromagnetic response of Ga$_{1-x}$Mn$_{x}$As across its phase diagram. We therefore provide a clear picture of the band structure of Ga$_{1-x}$Mn$_{x}$As through a detailed spectroscopic study of as-grown and annealed samples.  Recently post-growth annealing has enabled an increase in the $T_{C}$ and \textit{p}.\cite{annealpenn,annealnot} Annealing achieves this enhancement by removing Mn interstitials ($Mn_{i}$) acting as double donors and therefore compensating the holes.\cite{Mni,Mnsurface} Consistent with the notion of a boost in \textit{p}, after annealing the samples we find a large increase in the dissipative part of the optical conductivity ($\sigma_{1}(\omega)$) for all $\omega$ below the band gap of the GaAs host. However the overall shape of  $\sigma_{1}(\omega)$ in high-T$_C$ ferromagnetic films is remarkably similar to data from an earlier generation of samples.\cite{jasonpapers,DFH} The sum rule analysis of the electronic spectral weight allows us to discern the magnitude of the effective mass of the carriers ($m^{*}$), which is much larger than what has been predicted theoretically.\cite{sinova} Additionally $m^{*}$ is much larger than what is observed in p-type, non-magnetic GaAs doped to comparable levels,\cite{pgaas} where it is well established that $E_{F}$ lies in the GaAs VB. Furthermore we find a significant red-shifting of the mid-infrared resonance, which is also in direct contradiction to what is observed in p-type, non-magnetic GaAs,\cite{pgaas} and theoretical predications based on $E_{F}$ lying in the GaAs valence band.\cite{sinova} However similar qualitative behavior has been observed in n-type GaAs in the doping regime of a well defined impurity band.\cite{ngaas} Thus, we establish that the carriers reside in an IB at all values of \textit{p}. This conclusion is challenging to our current conception of the MIT in DMS, suggesting an important role for magnetism,\cite{Hellman} that had not been previously identified in doped semiconductors.

\begin{table}
\caption{\label{TBL}$T_{C}$ for the four Ga$_{1-x}$Mn$_{x}$As samples in this study, A indicates samples that have been annealed.}
\begin{ruledtabular}
\begin{tabular}{ccccc}
    x          &       0.052        &        0.073         &        0.052A        &        0.073A        \\
    \\
    $T_{C}~(K)$     &         80         &          80          &         120          &         140          \\
\end{tabular}
\end{ruledtabular}
\end{table}

    The samples were grown at UCSB on semi-insulating GaAs (100) by low temperature molecular beam epitaxy and annealed at PSU, see ref. \cite{annealpenn} for  details. After growth the wafers were split such that optical measurements could be performed on samples from the same growth both before and after annealing. The Ga$_{1-x}$Mn$_{x}$As layers had a nominal thickness of 40~nm to optimize the increase in $T_{C}$ upon annealing while still allowing for accurate optical measurements. All samples displayed a well defined hysteresis loop and $T_{C}$ when measured with a SQUID magnetometer (see table \ref{TBL}). Room temperature, ellipsometry between $0.62~eV-6~eV (5,000\rightarrow48,390~cm^{-1}$), at 75$^{\circ}$ angle of incidence as well as transmission over the range $0.005\rightarrow 1.42~eV$ ($40\rightarrow11,400~cm^{-1}$) from $292~\mathrm{K}$ to $7~\mathrm{K}$ were carried out at UCSD. Details of the measurements and extraction of optical constants is described in refs. \onlinecite{jasonpapers} $\&$ \onlinecite{kbelip}.

\begin{figure}
\includegraphics{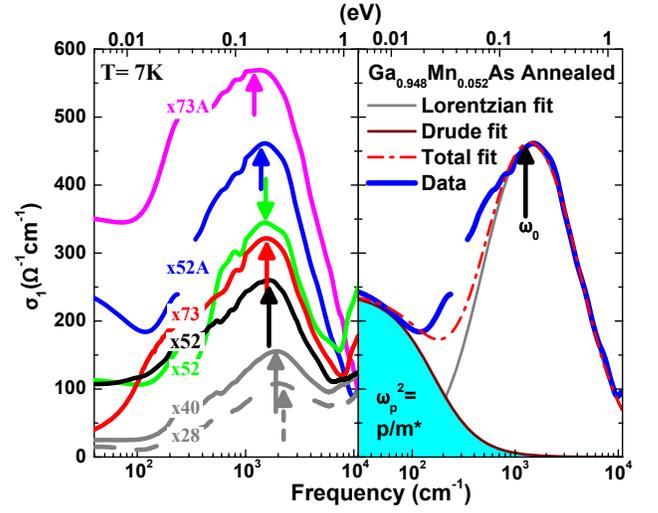}
\caption{\label{fig:s1comp} (color online) The real part of the conductivity for all samples in this study as well as the results of previous studies (Black and Grey).\cite{jasonpapers} A clear increase in $\sigma_{1}(\omega)$ is seen as the result of larger x and/or annealing. (Right) The results of fitting the 52A data with the two component model (eq. \ref{eq:sigma}).}
\end{figure}

We begin with an introduction to the semi-classical Drude-Lorentz model, which provides useful insights into the data. In this model one writes:
    \begin{equation}
    \label{eq:sigma} \sigma_{1}(\omega,x,T)=\frac{\Gamma_{D}^2\,{{\sigma }_{DC}}}
  {\Gamma_{D}^2 + {\omega }^2}+\frac{A\omega^{2}\Gamma_{L}}{(\Gamma_{L} \omega)^2 + (\omega^2 -\omega_{0}^{2})^2}
        \end{equation}
where the first term describes the response of free-carriers via the following parameters: $\Gamma_{D}$ the free carrier scattering rate, $\sigma_{DC}$ is the D.C. conductivity; and the second term describes the inter-band transition with $\omega_0$  its center frequency, $\Gamma_{L}$ its broadening and A its amplitude. One quantifies the strength of the free carrier response through the plasma frequency: $\omega_{p}^2=\frac{pe^2}{m^{*}} \cong \frac{2}{\pi}\int_{0}^{\Omega}\frac{\Gamma_{D}^2\,{{\sigma }_{DC}}}{\Gamma_{D}^2 + {\omega }^2}d\omega$, where e is the charge of the electron and $m^*$ the effective mass of the carriers.\cite{cutfoot} Therefore spectral weight in the far-infrared (FIR) is attributed to the free carriers (intra-band response)  and is proportional to \textit{p} divided by $m^{*}$. Examples of these shapes are shown on the right side of Fig. \ref{fig:s1comp} via a fit to the 52A data using  eq. \ref{eq:sigma}; wherein the distinct two-component character of the electronagetic response of annealed samples is uncovered.

\begin{figure}
\includegraphics{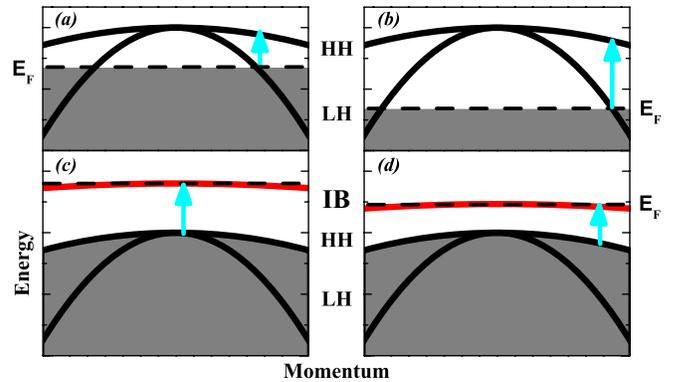}
\caption{\label{fig:bndstrc} (color online) Two scenarios for the electronic structure of Ga$_{1-x}$Mn$_{x}$As that have different implications for the resonant frequency of the MIR inter-band absorption. (a) $E_{F}$ lies in the light (LH) and heavy hole (HH) bands at low x, producing a transition between the two. (b) As \textit{p} is increased, $E_{F}$ moves further into the LH and HH bands, causing the transition to blue-shift. (c) If $E_{F}$ lies in the IB, then at low x a transition occurs from the VB to the IB. (d) At higher \textit{p}, $E_{F}$ moves deeper in the IB, causing the feature to red shift. }
\end{figure}

On the left side of Fig.  \ref{fig:s1comp}  we present $\sigma_{1}(\omega, x, 7K)$ for our new samples along with the spectra from our previous studies\cite{jasonpapers}. Some absorption is seen in the FIR in all samples, yet a clear Drude feature is only seen in the annealed films. In all of the films a resonance is observed in the mid-infrared (MIR) that by itself is consistent with both the VB and IB pictures of the electronic structure. Indeed, if $E_{F}$ lied in the GaAs VB, then a MIR resonance would result from transitions from the light to heavy hole bands.\cite{sinova} If on the other hand the holes reside in the IB, the MIR resonance results from transitions between the IB and VB.\cite{MillisDasSarma,jasonpapers,kbelip} We can discriminate between these scenarios by increasing \textit{p}. In particular the position of the MIR peak should either blue or red shift depending on the origin of the transition. A simple diagram of energy versus momentum (k) in Fig. \ref{fig:bndstrc} clarifies the rational for this assertion. The top two panels (a and b) assume the Mn induced IB has disolved into the VB, where $E_{F}$ now lies. In panel a, at low \textit{p}, an optical transition is realized between the light and heavy hole bands (LH and HH respectively). When \textit{p} increases,  $E_{F}$ moves deeper into the valence bands, (panel b) resulting in an optical transition that shifts to higher energies.\cite{sinova} If $E_{F}$ lies in a Mn induced impurity band (Mn$_{Ga}$) an optical transition between the LH, HH and Mn$_{Ga}$ bands will be observed. However, the separation between the IB and the VB is determined, in part, by the Coulomb attraction of the holes to the $Mn_{Ga}$. Thus as \textit{p} is increased (panels c and d), the Coulomb attraction is screened and the transition moves to lower $\omega$ as the IB moves closer to the VB.\cite{ngaas,MillisDasSarma}

\begin{figure}
\includegraphics{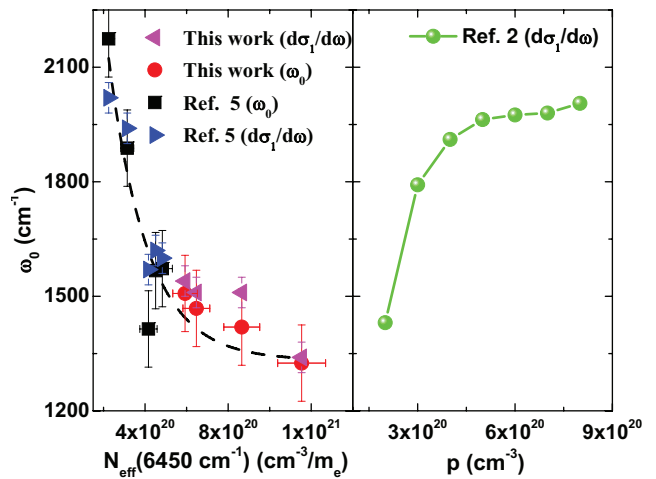}
\caption{\label{fig:wo} (color online)  Left panel: the peak position of the MIR feature determined by two alternative methods: setting $\frac{d\sigma_{1}(\omega,x)}{d\omega}=0$ and from the two-component analysis. (Left) This is plotted versus the spectral weight below $6,450~cm^{-1}$, which is proportional to the number of holes ($N_{eff}(6,450~cm^{-1})\propto{\textit{p}}$). This figure demonstrates that, regardless of the growth method, the MIR feature red shifts with increasing doping or \textit{p}. Right panel: the results from the predicted behavior of the GaAs VB versus \textit{p} for x=0.05, determined by the $\frac{d\sigma_{1}(\omega,x)}{d\omega}$ analysis.\cite{sinova}}
\end{figure}

To investigate these scenarios, we examined the position of the mid-IR resonance in the conductivity data  ($\omega_{0}$) using two complimentary methods. First, we looked for the maximum in $\sigma_{1}(\omega,x)$ by setting $\frac{d\sigma_{1}(\omega,x)}{d\omega}=0$. In addition, we have fit the data presented in Fig. \ref{fig:s1comp} with the two component model of Eq. \ref{eq:sigma} and obtained $\omega_{0}$  from the resonant energy in the second term of Eq. \ref{eq:sigma}.\cite{footnote1} The resulting center frequency of the inter-band transition is shown via arrows in Fig. \ref{fig:s1comp} and is plotted in Fig. \ref{fig:wo} as a function of the effective optical spectral weight, which is a measure of \textit{p}:
    \begin{equation}
    \label{eq:neff} N_{eff}(\omega_{c},x,T)=\frac{2}{\pi e^{2}}\int_{0}^{\omega_{c}}\sigma_{1}(\omega,x,T)d\omega \propto \frac{p}{m_{opt}}.
     \end{equation}
 with $\omega_{c}=6,450~cm^{-1}$.  This cutoff was chosen to provide a direct connection to the theoretical calculations of the optical properties, in particular the spectral weight of Ga$_{1-x}$Mn$_{x}$As.\cite{sinova} We have found that the increase in spectral weight from sample to sample is mostly independent of $\omega_{c}$, therefore the qualitative behavior shown in Fig. \ref{fig:wo} is not effected by the choice of cut-off. Finally, the results from setting $\frac{d\sigma_{1}(\omega,x)}{d\omega}=0$ and fitting the spectra using Eq. \ref{eq:sigma}, are in good quantitative agreement.

The data in Fig. \ref{fig:wo} demonstrate a rapid red-shift of the MIR feature with increase in \textit{p} as quantified via the magnitude of $N_{eff}$. Eventually $\omega_{0}$ levels off at approximately $1,350~cm^{-1}$. It is important to contrast these results with what has been seen in an optical study of p-type GaAs doped to similar levels, where $E_{F}$ clearly lies in the GaAs VB. In the latter system the inter-VB transistions clearly blue shift with the increase of \textit{p}.\cite{pgaas} The same qualitative trend is obtained in a recent theoretical work\cite{sinova}  aimed at the analysis of inter-VB absorption in Ga$_{1-x}$Mn$_{x}$As (see right panel of Fig.  \ref{fig:wo}). The position of the resonance plotted in this panel has been inferred from the model results for the conductivity by the $\frac{d\sigma_{1}(\omega,x)}{d\omega}$ method. Therefore our observation of red-shiftiing of $\omega_{0}$ is in direct contradiction to both experiemntal results and theoretical modeling of the optical response due to E$_F$ situated in the VB of GaAs. Furthermore the prediction of ref. \onlinecite{sinova} for the magnitude of $\omega_{0}$ is too high in energy to be consistent with our data.

Next we inquire into the nature of the states at $E_{F}$ through the examination of the plasma frequency. Since we are able to separate out the free carrier component in the annealed films, this analysis is particularly fruitful as we can then determine the effective mass of the holes: $m^{*}\propto\frac{p}{\omega_{p}^{2}}$. Previous optical studies of p-type GaAs where $E_{F}$ lies in the VB have demonstrated that $m^*=0.38~m_{e}$.\cite{pgaas} Assuming between 0$\%$ and 50$\%$ compensation of the Mn for 52A (73A) sample we find $36~m_{e}<m^*<72~m_{e}$ ($16~m_{e}<m^*<33~m_{e}$). Furthermore, theoretical calculations suggest that if $E_{F}$ lies in the VB, then an optical mass ($m^{opt}\propto \frac{p}{N_{eff}(800~meV)}$) should be independent of \textit{p} and lie between $0.25~m_{e}\rightarrow0.29~m_{e}$.\cite{sinova} Using the same assumptions for \textit{p} we find $0.7~m_{e}<m^{opt}<1.4~m_{e}$ for both the 52A and 73A films. The $m^{*}$ reported here likely reflect strong interactions in Ga$_{1-x}$Mn$_{x}$As, yet the carriers are still too heavy for $E_{F}$ to lie in the GaAs VB.

Since it is well established that the disorder in Ga$_{1-x}$Mn$_{x}$As is strong and is likely to become more significant as x is increased (due to the additional $Mn_{i}$ that are introduced)\cite{Mni}, it is important to discuss what effects impurities may have on our results. One may be tempted to hypothesize that the red-shifting results from the additional defects in the samples. However annealing reduces the disorder and enhances the carrier density by removing $Mn_{i}$. In fact, our observation of a clear Drude feature only in annealed samples, see Fig. \ref{fig:s1comp}, confirms a \textit{reduction} of scattering upon annealing the films. Nonetheless, the MIR feature continues to red shift after annealing, suggesting that disorder plays little to no role in it's position. This is not surprising, since our discussion of the red-shifting of $\omega_{0}$ with \textit{p} does not rely on k conservation, only on the energy difference between the center of the IB and the top of the VB. Furthermore, disorder generally broadens the observed width of optical transitions, but does not significantly affect their position. Finally, as discussed below, the clear observation of critical points in Ga$_{1-x}$Mn$_{x}$As\cite{kbelip, XMCD} argues against disorder significantly effecting the optical properties of these samples. 

The data in Figs. \ref{fig:s1comp} and \ref{fig:wo} provides conclusive evidence that the holes in Ga$_{1-x}$Mn$_{x}$As resides in a Mn induced impurity band regardless of carrier density, thereby establishing the basic model of the electronic structure of this prototypical magnetic semiconductor. This conclusion has a number of interesting implications and raises some important new questions.  First, unmistakable evidence for IB conduction in Ga$_{1-x}$Mn$_{x}$As suggests that heavy masses associated with states at $E_{F}$ have to be taken into consideration in designing device functionalities involving spin and/or charge injection as well as magneto-optical effects. Second, the large values of $m^*$ we have found provides an explanation of the rather low mobility ($\mu=\frac{e\tau}{m^*}$) of Ga$_{1-x}$Mn$_{x}$As. Intriguingly, even the cleanest samples of Ga$_{1-x}$Mn$_{x}$As demonstrating high values of T$_{C}$ reveal $\mu$ as low as $1-5~cm^2/Vs$, which is one to two orders of magnitude smaller than in GaAs doped to similar concentrations with non-magnetic impurities. In fact, to produce such low mobility's in III-V materials, one generally has to make the crystals amorphous. However all films in this study reveal Van-Hove singularities via spectroscopic ellipsometry\cite{kbelip} and optical magneto circular dichroism (MCD)\cite{XMCD}, indicating that the carrier momentum is still a good quantum number. One can also evaluate the strength of disorder via a calculation of the product of the mean free path (\textit{l}) and fermi momentum ($k_{F}$). Assuming a single band and the same values of \textit{p} used for the extraction of the effective mass, in the annealed samples we find: $3<k_{F}\textit{l}<5$. Since these values are significantly larger than one they indicate that the transport in these samples is coherent. Therefore the low mobility of Ga$_{1-x}$Mn$_{x}$As cannot originate solely from low values of $\tau$. Thus the low values of $\mu$ cannot be understood as simply arising from disorder, implying that heavy effective masses reported here are the primary cause of the small mobility in Ga$_{1-x}$Mn$_{x}$As. 

	Interestingly recent optical and X-ray MCD studies\cite{XMCD} have suggested the exchange constant is rather large in Ga$_{1-x}$Mn$_{x}$As, which may explain the significant enhancement of $\frac{m^{*}}{m_{e}}$ that we observe. Specifically it has been shown by numerous authors that large values of the exchange will tend to localize the holes around the Mn.\cite{imthry, MillisDasSarma, SamarthReview} Furthermore, the localization effect described above may also account for the persistence of the IB in annealed Ga$_{1-x}$Mn$_{x}$As.\cite{imthry, MillisDasSarma} Nonetheless the existence of the IB presents a significant challenge to our current conception of the MIT in doped semiconductors. In particular, it is a long held belief that the IB that emerges at low doping levels is built purely from hydrogenic states of the acceptor, and therefore can never produce Bloch waves. The metallic transport in doped semiconductors is then understood via the assumption that once the Coulomb attraction between the holes and the acceptors is completely screened the IB "dissolves" into the main band. This then implies that the holes now occupy Bloch states, resulting in metallic behavior. Our results suggest that this picture is incomplete when doping is accomplished with magnetic impurities. Interestingly recent tight-binding calculations suggest a Mn-induced resonance in the VB, which may account for our results.\cite{Flatte} Specifically, while our results require the presence of an IB, they cannot exclude its' overlap with the VB in the density of states. Clearly further studies are needed to clarify the interplay between the carrier dynamics, band structure, and ferromagnetism of Ga$_{1-x}$Mn$_{x}$As. 

 Work at UCSD was supported by the DOE and NSF, and the work at UCSB and PSU were supported by DARPA and ONR. We are grateful for our discussions with L. Cywinski, M. E. Flatte, A. MacDonald, A.J. Millis, S. Das Sarma, J. Sinova, J.-M. Tang, and C. Timm.


\begin{thebibliography}
\bibitem{}
\bibitem{SamarthReview} A.H. MacDonald et al., Nature Materials \textbf{4}, 195 (2005).
\bibitem{sinova} J. Sinova et al., Phys Rev. B. \textbf{66}, 041202 (2002).
\bibitem{vbthry}  B. Lee et al., Semicond. Sci. Technol. \textbf{17}, 393 (2002); T. Dietl et al., Science \textbf{287}, 1019 (2000).
\bibitem{imthry} M. Berciu, and R.N. Bhatt, Phys. Rev. B \textbf{69} 045202 (2004); G. Alvarez and E. Dagotto, Phys. Rev. B \textbf{68}, 045202 (2003).S. Sanvito et. al., Phys. Rev. B. \textbf{63} 165206 (2001); S. C. Erwin and A.G. Petukhov, PHys. Rev. Lett. \textbf{89}, 227201 (2002); P. Mahadevan and A. Zunger, Phys. Rev. B. \textbf{69}, 115211 (2004); M.A. Majidi et al., cond-mat/0510716.
\bibitem{MillisDasSarma} E.H. Hwang et al., Phys. Rev. B \textbf{65}, 233206 (2002).
\bibitem{Flatte}J.M. Tang et al., Phys. Rev. Lett. \textbf{92}, 047201 (2004).
\bibitem{moreno} J. Moreno et al., Phys. Rev. Lett. \textbf{96}, 237204 (2006).
\bibitem{jasonpapers}E.J Singley et al., Phys Rev. Lett. \textbf{89}, 097203 (2002); E.J Singley et al., Phys Rev. B. \textbf{68}, 165204 (2003).
\bibitem{kbelip} K. S. Burch et al., Phys. Rev. B. \textbf{70}, 205208 (2004).
\bibitem{DFH} K. S. Burch et al., Phys. Rev. B \textbf{71}, 125340 (2005).
\bibitem{Fujimori} O. Rader et. al., Phys. Rev. B \textbf{69}, 075202 (2004); J. Okabayashi et. al., Physica E \textbf{10}, 192 (2001).
\bibitem{Ploog}V. F. Sapega et al., Phys. Rev. Lett. \textbf{94} 137401 (2005).
\bibitem{annealpenn}S.J. Potashnik et al., Appl. Phys. Lett. \textbf{79}, 1495 (2001);
\bibitem{annealnot}K.W. Edmonds et al., Appl Phys. Lett. \textbf{81}, 3010 (2002); T. Hayashi et al., Appl. Phys. Lett \textbf{78}, 1691 (2001).
\bibitem{Mni}K.W. Edmonds et al., Phys. Rev. Lett \textbf{92}, 37201(2004); K. M. Yu et al., Phys. Rev. B \textbf{65}, 201303 (R) (2002).
\bibitem{Mnsurface}M. B. Stone et al., Appl. Phys. Lett. \textbf{83}, 4568 (2003); W. Limmer et al., Phys. Rev. B \textbf{71}, 205213 (2005); V. Stanciu et al., cond-mat/0505040.
\bibitem{Hellman} D.N. Basov et al., Europhys. Lett. \textbf{57}, 240 (2002).
\bibitem{cutfoot} We note that this sum-rule assumes the free-carrier response is seperated from the inter-band response, and that the cuttoff ($\Omega$) is choosen to be many times $\Gamma_{D}$.
\bibitem{ngaas}S. Liu et al., Phys. Rev. B \textbf{48}, 11394 (1993).
\bibitem{footnote1} We set $\sigma_{DC}=\sigma(30~cm^{-1})$ to reduce the number of fit parameters.
\bibitem{fitfoot}Fitting the results of ref. \cite{sinova} with the two component model indicates that their $\omega_{0}$ is approximately 500 $cm^{-1}$ greater than that displayed with green dots in Fig. \ref{fig:wo}.
\bibitem{pgaas} W. Songprakob et. al., J. Appl. Phys. \textbf{91}, 171 (2002).
\bibitem{XMCD} D. J. Keavney et al., Phys. Rev. Lett \textbf{91}, 187203 (2003);K. Ando et al., J. Appl. Phys. \textbf{83}, 6548 (1998); B. Beschoten et al., Phys. Rev. Lett \textbf{83}, 3073 (1999).
\end{thebibliography}
\end{document}